\documentclass[article,ij4uq]{ij4uq}              

\usepackage[hang]{footmisc}
\fancypagestyle{plain}{%
  \fancyhf{}
  \fancyhead[R]{\small {\it International Journal for Uncertainty Quantification}, x(x): \thepage--\pageref{LastPage} (\myyear\today)}
  \fancyfoot[R]{\small\bf\thepage }
  \fancyfoot[L]{\fottitle}
  }
\renewcommand{\dmyy}{2020}
\renewcommand{\myyear}{2020}
\renewcommand{\today}{}
\def\beq{\begin{equation}}
\def\eeq{\end{equation}}
\def \vect#1{\bm{#1}}

\begin{document}

\volume{Volume x, Issue x, \myyear\today}
\title{Multi-fidelity estimators for coronary circulation models under clinically-informed data uncertainty}
\titlehead{Multi-fidelity UQ on coronary artery simulation}
\authorhead{Jongmin Seo, Casey Fleeter, Andrew M. Kahn, Alison L. Marsden, \&  Daniele E. Schiavazzi}
\author[1]{Jongmin Seo}
\author[2]{Casey Fleeter}
\author[3]{Andrew M. Kahn}
\author[1]{Alison L. Marsden}
\corrauthor[4]{Daniele E. Schiavazzi}
\corremail{dschiavazzi@nd.edu}
\corraddress{Department of Applied and Computational Mathematics and Statistics, University of Notre Dame, IN, USA}
\address[1]{Department of Pediatrics (Cardiology), Bioengineering and ICME, Stanford University, Stanford, CA, USA}
\address[2]{Institue for Computational Mathematics and Engineering, Stanford University, Stanford, CA, USA}
\address[3]{Department of Medicine, University of California San Diego, La Jolla, CA, USA}
\address[4]{Department of Applied and Computational Mathematics and Statistics, University of Notre Dame, IN, USA}


\dataO{11/15/2019}
\dataF{mm/dd/2020}

\abstract{Numerical models are increasingly used for non-invasive diagnosis and treatment planning in coronary artery disease, where service-based technologies have proven successful in identifying hemodynamically significant and hence potentially dangerous vascular anomalies.
Despite recent progress towards clinical adoption, many results in the field are still based on a deterministic characterization of blood flow, with no quantitative assessment of the variability of simulation outputs due to uncertainty from multiple sources. 
In this study, we focus on parameters that are essential to construct accurate patient-specific representations of the coronary circulation, such as aortic pressure waveform, intramyocardial pressure and quantify how their uncertainty affects clinically relevant model outputs.
We construct a deformable model of the left coronary artery subject to a prescribed inlet pressure and with open-loop outlet boundary conditions, treating fluid-structure interaction through an Arbitrary-Lagrangian-Eulerian framework.
Random input uncertainty is estimated directly from repeated clinical measurements from intra-coronary catheterization and complemented by literature data.
We also achieve significant computational cost reductions in uncertainty propagation thanks to multifidelity Monte Carlo estimators of the outputs of interest, leveraging the ability to generate, at practically no cost, one- and zero-dimensional low-fidelity representations of left coronary artery flow, with appropriate boundary conditions. 
The results demonstrate how the use of multi-fidelity control variate estimators leads to significant reductions in variance and accuracy improvements with respect to traditional Monte-Carlo. In particular, the combination of three- dimensional hemodynamics simulations and zero-dimensional lumped parameter network models produces the best results, with only a negligible (less than one percent) computational overhead.
}

\keywords{Cardiovascular simulation, Multi-fidelity framework, Coronary artery hemodynamics, Uncertainty quantification.}

\maketitle

\section{Introduction}

Cardiovascular disease poses significant burden on the lives of millions of people worldwide, with projected total costs estimated by the American Heart Association of over 1 trillion dollars by 2035~\cite{AHA2018}.
As the standard of care improves through continued research and innovation, \emph{virtual} hemodynamic representations have been proposed as a non-invasive, image-based approach to complement clinical diagnostics, assess patient risk, aid in clinical decision-making and facilitate treatment planning.
Use of numerical models in clinical diagnosis for specific pathologies has also been extensively tested and supported by clinical studies ~\cite{koo2011diagnosis, min2012diagnostic, norgaard2014diagnostic}. 
The complexity of these models has substantially increased in recent years, transitioning from idealized and patient-agnostic concepts to patient specific models constructed through a complex pipeline which includes image acquisition and segmentation, application of physiologically sound boundary conditions, specification of heterogeneous wall material properties and accurate solution of the governing equations on high-performance computing platforms~\cite{Marsden2014, Marsden2015}. { Our cardiovascular simulation methods have been extensively validated against {\it in vitro} data \cite{Kung2011b, Kung2014,Steinman2012}, and have been applied to clinical decision making including non-invasive assessment of fractional flow reserve (FFR)~\cite{taylor2013computational}, new surgical designs for congenital heart disease~\cite{yang2010constrained}, risk assessment in coronary artery bypass surgery~\cite{Ramachandra2017}, thrombotic risk stratification in Kawasaki disease~\cite{Noelia2017}, design of ventricular assist devices~\cite{Long2014}, study of cerebral and aortic aneurysms~\cite{Humphrey2008}, stent design and placement~\cite{migliavacca2002mechanical, Gundert2012} and many others.}

Although the realism of simulation tools has improved in recent years, the predictions provided through these tools are essentially deterministic in most of the literature. Most cardiovascular simulations ignore impact of variability due to uncertainty in input parameters on simulation outcomes. In our view, this shortcoming can be overcome through the establishment of strict guidelines and effective methods to assess the impact of uncertainty on simulation predictions, particularly for model-as-a-device technologies used in the clinic.
In particular, there is a need for methods that can increase efficiency of uncertainty quantification to make it tractable for full-scale patient-specific problems.  Common sources of uncertainty relate to basic hemodynamic metrics (e.g., heart rate or blood pressure), echocardiography measures(stroke volume, ejection fraction, cardiac output, acceleration times) and invasive cardiac catheterization data (cardiac pressures or intravascular ultrasound velocities). Inter-patient variability from population studies as well as intra-patient variability from repeated measurements, can also be useful in complementing information on uncertainty.

After characterizing the input variability, output statistics are determined by \emph{uncertainty propagation} through a computationally expensive three-dimensional cardiovascular model. This step can easily become prohibitive, particularly for models with large discretizations, models that account for fluid-structure interaction~\cite{Figueroa2006} or include physiologic boundary conditions~\cite{Esmaily2013res}, possibly assimilated from available clinical data under uncertainty~\cite{Tran2017, Schiavazzi2017}.
Several studies in the literature have investigated the effects of parametric uncertainty in the context of coronary artery disease, for example considering one-dimensional hemodynamic models and the effect of variability in constitutive model parameters~\cite{Xiu2007}, arterial wall stiffness, inlet velocity~\cite{Brault2016}, 
combined boundary conditions and material properties~\cite{Tran2019},
physiologic and anatomic parameters that impact the predictions of FFR~\cite{yin2019one}
resistance and pressure~\cite{Chen2013}, and assessment of global sensitivity.
In most of these studies, uncertainty propagation is performed using spectral stochastic approaches based on generalized polynomial chaos expansion~\cite{xiu2002wiener}, multi-element~\cite{wan2005adaptive} or multi-resolution~\cite{le2004multi, schiavazzi2014sparse, schiavazzi2017generalized} approaches, 
probabilistic tessellations~\cite{witteveen2012simplex}, compressive sampling and LASSO~\cite{doostan2011non,blatman2011adaptive} and others.
While offering significant computational savings when the problem at hand has a smooth stochastic response under a moderate dimensionality, these methods suffer significantly from the increase in computational complexity of handling tensor product basis in high dimensions.

Acceleration, more precisely \emph{variance reduction}, in Monte Carlo sampling has been widely discussed in the literature but, more recently, new multifidelity Monte Carlo estimators are increasingly appearing in applications (see a recent review in~\cite{peherstorfer2018survey}). In this study, we focus on approximate control variate estimators~\cite{gorodetsky2018generalized}. 
The basic idea is to shift the computational cost from the solution of an expensive high-fidelity model to that of a family of inexpensive low-fidelity surrogates, with the only requirement for high- and low-fidelity models being that they are sufficiently correlated. In other words, the low fidelity models do not necessarily need to be an exact approximation of higher fidelity models, as long as they are correlated. 
Our group has recently demonstrated the efficiency of such estimators on fully pulsatile simulations using healthy and diseased cardiovascular geometries~\cite{fleeter2019multilevel}, but further applications of multi-fidelity approach research remains to be conducted in cardiovascular simulations.
In this study, we focus on a vascular submodel with boundary conditions that are specifically designed to capture coronary physiology~\cite{kim2010patient,sankaran2012patient}. We aim to quantify the variability of clinically relevant model outputs, based on uncertainty in input parameters that are often subjectively chosen by the analyst. The following original contributions are discussed in this paper:

\vspace{-7pt}

\begin{itemize}
\setlength\itemsep{0pt}
\item Random input uncertainty is informed directly through repeated measurements from intra-coronary chatheterization. To the authors knowledge, this is one of the few studies where clinical data collection is designed to provide sufficient data so that uncertain random inputs can be statistically characterized.
\item We focus on the most important parameters, whose characterization is essential for accurately simulating the coronary arterial circulation in specific patients. For example, the intra-myocardial pressure and its time derivative are difficult to determine in practice, and often assumed equal to the left ventricular pressure.
\item Coronary boundary conditions have been implemented for all model fidelities, in particular for one- and zero-dimensional flow solvers.
\item We demonstrate the effectiveness of multi-fidelity approaches in models incorporating wall deformation with Arbitrary-Lagrangian-Eulerian (ALE) framework.  
\end{itemize}

\vspace{-3pt}

\noindent The paper is organized as follows.
Section~\ref{sec:LCAmodeling} presents the formulation for the three-, one- and zero-dimensional hemodynamic solvers used to analyze the coronary circulation and discusses the boundary conditions used to mimic the coronary physiology. 
This is followed by a discussion of approximate control variate estimators and their variance in Section~\ref{sec:MFUQ}.
Section~\ref{sec:UQmodeling} focuses on the random inputs and their representation in terms of a random combination of time-varying modes. We also identify the quantities of interest, the computational cost needed for their determination and the correlations across different model fidelities.
Variance reduction and accuracy for the selected Monte Carlo estimators are analyzed in Section~\ref{sec:results} and finally, in Section~\ref{sec:Discussion}, we summarize our findings, discuss limitations and briefly outline future work.

\section{Multi-dimensional models for the coronary circulation}\label{sec:LCAmodeling}
\begin{figure}[ht!]
\vspace{-6pt}
\centering
\includegraphics[width=0.5\textwidth, keepaspectratio]{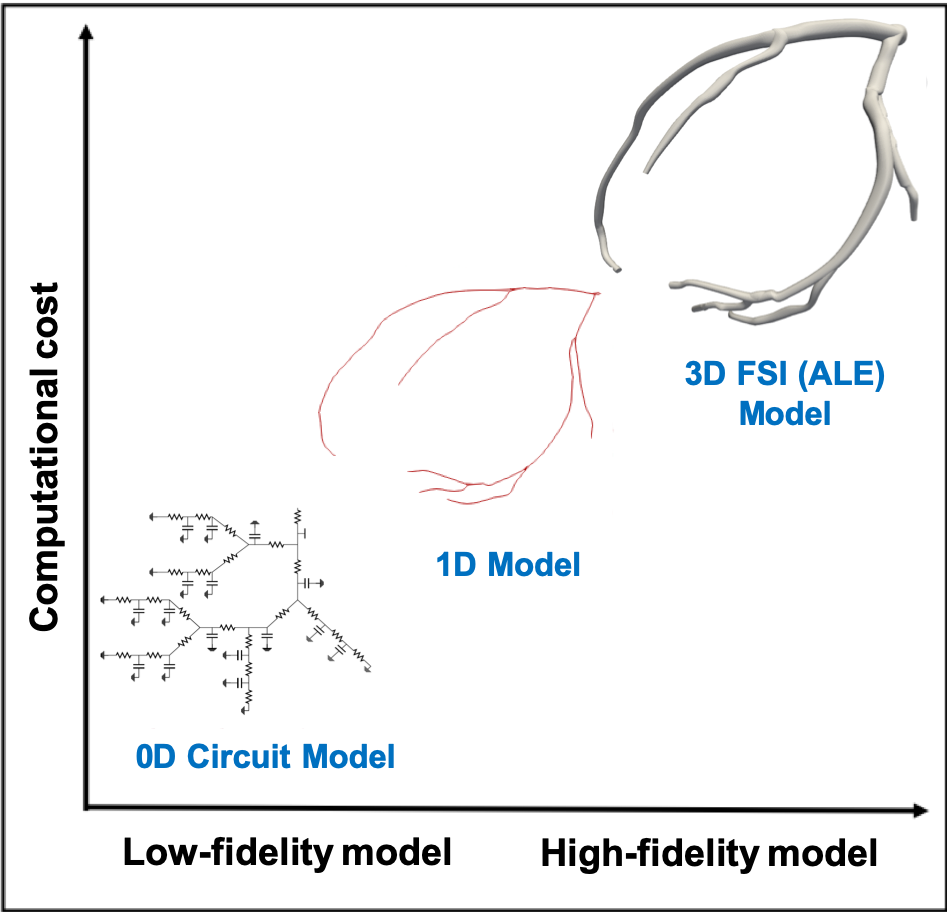}
\caption{Computational cost versus fidelity for the three hemodynamic models utilized in this study.}\label{fig:1}
\end{figure}

\begin{figure}[ht!]
\vspace{-6pt}
\centering
\includegraphics[width=1.0\textwidth, keepaspectratio]{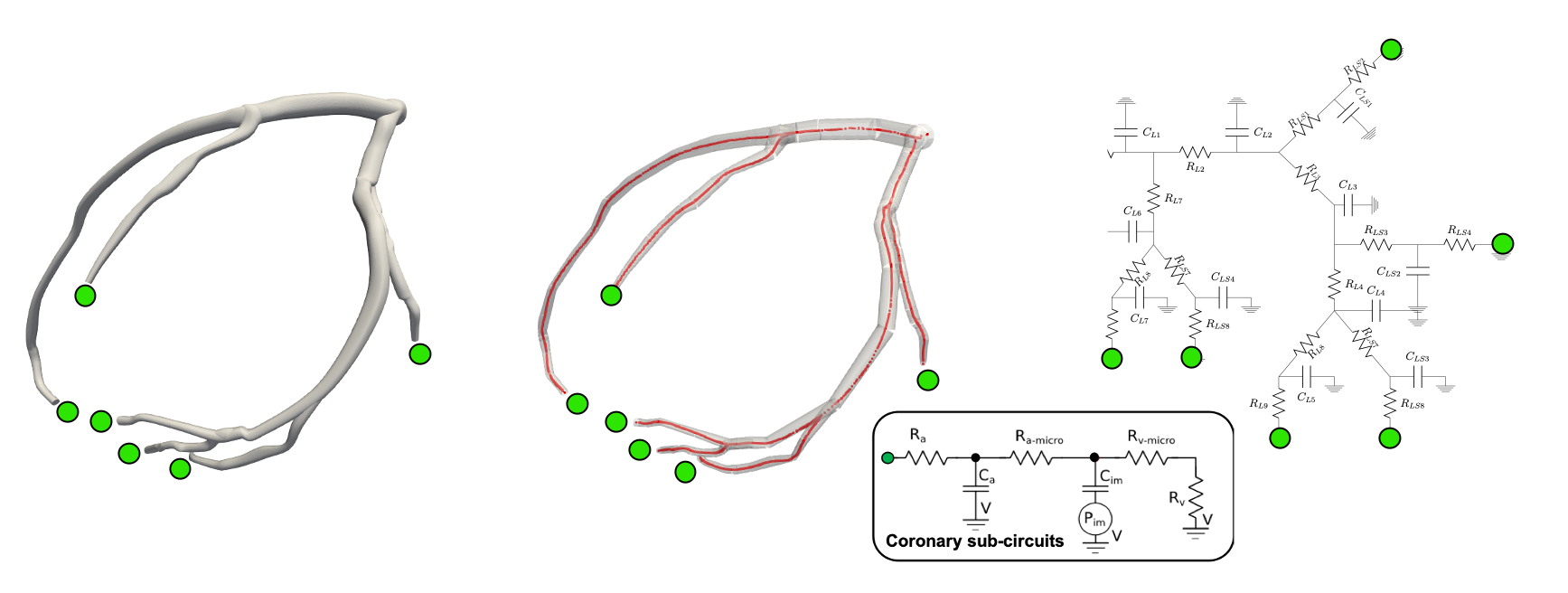}
\caption{Numerical models with coronary boundary conditions at the outlets. Three-dimensional model with deformable walls (left), one-dimensional model (center) and zero-dimensional lumped parameter network model (right).}\label{fig:2}
\end{figure}

\subsection{Three-dimensional left coronary model}

A three-dimensional anatomic model of the left coronary artery was constructed from Computed Tomography clinical images using SimVascular~\cite{Updergrove2016}. 
Outlet boundary conditions are specified through a lumped parameter network (LPN) representation of the downstream circulation whose parameters are tuned to match multiple clinical targets (stroke volume, ejection fraction, pressure etc.) and found to reproduce physiologically admissible responses~\cite{Tran2017, Seo2019}. 
Interaction between fluid and structure is simulated through Arbitrary Lagrangian-Eulerian (ALE) coupling provided through the SimVascular svFSI solver, which implements a variational multi-scale finite element method with second order implicit generalized-$\alpha$ time integration~\cite{Bazilevs2007, Esmaily2015}. 
The incompressible Navier-Stokes equations in ALE form are
\begin{equation}
\begin{split}
\rho\,\frac{\partial \mathbf{u}}{\partial t}|_{\hat{ \mathbf{x}}}+\rho\,{\mathbf{v}}\cdot\nabla\,{\mathbf{u}}  &= \rho\,\mathbf{f}+ \nabla \cdot {\mathbf{\sigma}_f}\\
\nabla  \cdot {\mathbf{u}} &= 0
\end{split}
\quad\text{in}\,\,\Omega_f,
\end{equation}
where $\rho, \mathbf{u}=\mathbf{u}(\mathbf{x},t)$, and $\mathbf{f}$ are fluid density, velocity vector, and body force in the fluid domain $\Omega_f$.
Blood is assumed as a Newtonian fluid, for which $\mathbf{\sigma}_f=-p\,{\mathbf{I}} + \mu\,(\nabla{\mathbf{u}}+\nabla{\mathbf{u}}^T) = -p\,{\mathbf{I}} + \mu\,\nabla^{s}{\mathbf{u}}$, with viscosity $\mu$, pressure $p=p(\mathbf{x},t)$, and $\mathbf{v}=\mathbf{u-\hat{u}}$ is the fluid velocity relative to the mesh.
Additionally, in the solid domain $\Omega_s$, we solve the equilibrium equations
\begin{equation}
\begin{split}
\rho_s\,{\frac{\partial \mathbf{u}}{\partial t}}=\rho_s\,\mathbf{f} + \nabla\cdot \boldsymbol{\sigma}_s
\end{split}
\quad\text{in}\,\,\Omega_s,
\end{equation}
where $\rho_s$ and $\boldsymbol{\sigma}_s$ denote the density and stress tensor in the solid, respectively.
The spatial discretization is based on the variational multi-scale finite element method \cite{Bazilevs2007, Esmaily2015, Seo2019} and P1-P1 (linear and continuous) elements with co-located nodal velocity and pressure unknowns, while a Saint Venant-Kirchhoff hyper-elastic constitutive model is used for the solid wall. For a complete formulation of the above equations in weak form and the resulting algebraic system, the interested reader is referred to~\cite{Esmaily2015, Seo2019}. 
Linear system solutions are computed using the Trilinos library~\cite{Trilinos}, developed at Sandia National Laboratory and coupled with the SimVascular svFSI solver. We use either the Bi-Conjugate Gradient iterative linear solver with incomplete LU preconditioner or the Generalized Minimum Residual with a diagonal preconditioner. These combinations were shown to be optimal for cardiovascular simulations with deformable walls in our prior work~\cite{Seo2019}.
The computational mesh for the coronary artery lumen was informed by a preliminary convergence study~\cite{Seo2020} and contains 567,373 tetrahedral elements. The wall mesh consists of 373,435 tetrahedral elements, with three elements through the thickness, which is regarded as appropriate due to the prevalent membrane deformations.

\subsection{One-dimensional left coronary model}

The formulation of the one-dimensional hemodynamics solver used in this study, and available through the SimVascular project, is adapted from Hughes and Lubliner \cite{Hughes1973}, with details discussed in \cite{Wan2002, Vignon2004, Vignon2006}. 
Blood is assumed Newtonian, flowing in the axial direction ($z$) of an ideal cylindrical branch, the pressure is assumed constant over the entire vessel cross section, and no-slip boundary conditions are applied at the lumen. 

Conservation of mass and momentum are formulated by integrating the incompressible Navier-Stokes equations over the cross section of a deformable cylindrical domain as
\begin{equation}
\frac{\partial A}{\partial t}+\frac{\partial Q}{\partial z} = 0,\;\;\; \frac{\partial Q}{\partial t}+\frac{\partial }{\partial z}\left[(1+\delta)\frac{\partial Q^2}{\partial A}\right]+\frac{A}{\rho}\frac{\partial \rho}{\partial z}=Af +N\frac{Q}{A} +\nu\frac{\partial^2 Q}{\partial z^2}. 
\label{eq:1DM}
\end{equation}
Here the solution variables are the vessel cross-sectional area $A$ and blood flow rate $Q$; other parameters are the density $\rho$, external force $f$, kinematic viscosity $\nu=\mu/$, and velocity profile parameters $\delta$ and $N$ defined as
\begin{equation}
\delta = \frac{1}{A} \int_{A} (\phi^2-1)ds,\;\;\;N=\nu \int_{\partial A} \frac{\partial \phi}{\partial m} dl, 
\label{eq:1Daddtional}
\end{equation}
{ where $\phi(r)$ is a general velocity profile function given in the form $v(z)=\phi(r) \bar{v}$ with average velocity $\bar{v}$ and radius $r$.}
A linear constitutive equation is given by 
\begin{equation}
\bar{p}(A,z)=p_0(z)+\left(\frac{Eh_0}{r_0(z)}\right)\left(\sqrt{\frac{A(z,t)}{A_0(z)}}-1\right),
\label{eq:1DC}
\end{equation}
where $E$ is the Young's modulus of the vascular tissue, $h_0$ the wall thickness, $r_0(z)$ and $A_0(z)$ the undeformed inner radius and area, respectively.
Our in-house 1D solver combines a stabilized finite element discretization in space and a discontinuous Galerkin approach in time; the non-linear algebraic system is solved using modified Newton iterations, while pressure and cross-sectional area continuity at the joints between two segments is enforced through Lagrange multipliers.
Location and properties of joints, segments, boundary conditions, data tables, initial conditions and solver parameters are specified through a text input file. Input flow and pressure waveforms are prescribed through time/frequency data tables, while supported boundary conditions include pressure, area, flow, resistance (steady or time-varying), pressure wave, RCR, coronary, impedance and admittance. Further details can be found in~\cite{Wan2002}. 

\subsection{Zero-dimensional left coronary model}

Our simplest low-fidelity representation consists of a lumped parameter network, an equivalent circuit layout formulated by hydrodynamic analogy, with flow rate and pressure as the main unknowns. Each circuit element is associated with an algebraic or differential equation, so that
\begin{equation}
\text{resistor}\;\Delta P =RQ,\;\text{capacitor}\;\Delta Q=C \frac{d P}{dt},\;\text{inductor}\;\Delta P =L \frac{dQ}{dt}.
\end{equation} 
Resistors, capacitors and inductors are used to represent viscous dissipation in vessels through friction at the endothelium, vascular tissue compliance and blood inertia, respectively.
A Poiseuille flow assumption is used to determine the model parameters for these circuit elements \cite{Milisic2004},
\begin{equation}
R=\frac{8\mu l }{\pi r^4},\;\; C=\frac{3 l \pi r^3}{2Eh},\;\;L=\frac{l \rho}{\pi r^2},
\end{equation} 
where $\mu$ is the dynamic viscosity, $l$ is the vessel length, $r$ is the vessel radius, $E$ is the elastic modulus, $h$ is the wall thickness and $\rho$ is the blood density.
An in-house automated 0-D solver is used in this study, designed to formulate arbitrary circuit layouts from the integration of simple templates, such as capacitors, inductors, resistors, junctions, pressure sources, flow sources, and diodes. Each template is connected via edges and junctions with user-specified connectivity among the templates. The equations are then automatically assembled in a system of differential-algebraic equations,
\beq
E(y,t)\,\dot{y}+F(y,t)+C(t)=0, 
\eeq
which is solved numerically. The above equation is discretized in time and integrated in time using an implicit, generalized-$\alpha$ scheme.

\subsection{LPN boundary conditions at the coronary arteries}

Specialized boundary conditions are applied in this study to three-, one- and zero-dimensional model fidelities, to better capture the physiology of coronary blood flow~\cite{Kim2009}.
In this context, it is well known how the contraction of the cardiac muscle during systole impedes the flow in the coronaries, which instead reaches its maximum following diastolic relaxation. In other words, a typical coronary flow waveform is out-of-phase with the aortic pressure (Figure \ref{fig:3}).
To mimic this behavior, a special \emph{coronary} boundary condition has been proposed in the literature~\cite{Kim2009}, consisting of an RCRCR circuit connected to an intramyocardial pressure generator.
Coronary boundary conditions~\cite{Kim2009, Sankaran2012} are applied to the $n_{o}=6$ left coronary artery (LCA) outlets, formulated through the ordinary differential equations
\begin{equation}
\frac{dP_\text{p,i}}{dt}=\frac{1}{C_\text{a,i}}\Big(Q_{i}-\frac{P_\text{p,i}-P_\text{d,i}}{R_\text{am,i}}\Big),\;\;i=1,2,...,n_o, 
\end{equation}
\begin{equation}
\frac{dP_\text{d,i}}{dt}=\frac{1}{C_\text{im,i}}\Big(\frac{P_\text{p,i}-P_\text{d,i}}{R_\text{am,i}}-\frac{P_\text{d,i}-P_\text{im}}{R_\text{v,i}}\Big)+\frac{dP_\text{im}}{dt}, \;\;i=1,2,...,n_o, 
\end{equation}
where $P_\text{p,i}$ and $P_\text{d,i}$ are the proximal and distal pressures, $C_\text{a,i}$ and $C_\text{im,i}$ are the proximal and distal capacitances and $R_\text{am,i}$ and $R_\text{v,i}$ are the resistances, respectively (Figure~\ref{fig:2}). 
Note that the same rate of intramyocardial pressure $dP_\text{im}/dt$ is used for all $n_o$ outlets.
In the 3D model, the above equations are integrated in time using a fourth-order Runge-Kutta explicit scheme, while the pressure at the 3D model outlet $P_\text{o,i}$ is computed as $P_\text{o,i}=P_\text{p,i}+R_\text{a,i}\,Q_\text{i}$ and coupled to the three-dimensional model solution at each time step \cite{Esmaily2012}.
In the 1D model, algebraic equations of the coronary circuit are instead implemented as discussed in~\cite{Kim2009}, whereas the compartment containing the coronary boundary conditions simply extends the 0D circuit network and these two are solved in a monolithic fashion. 

The total coronary resistance was computed by assigning 4$\%$ of the cardiac output to the coronary arteries~\cite{Bogren1989} and vessel resistances were distributed among outlets following a morphometric relation associating flow rates with vessel diameters, $Q\propto (d/2)^{m}$. 
An empirically derived morphometry exponent $m=2.6$ has been suggested in~\cite{Zhou1999,Changizi2000}; for the coronary circulation, a previous uncertainty quantification study found only negligible difference in the simulation results with $m$ ranging between 2.4 and 2.8~\cite{Seo2020}.
Approximating the diameter with the square root of the area $\sqrt{A_{i}}$, the resistance of a distal branch, $R_{i}$ is computed as 
\begin{equation}
R_i=\frac{\sum_{j} \sqrt{A_j^{2.6}}}{\sqrt{A_i^{2.6}}}\cdot R_\text{total},\;\;\;i,j=1,2,...,n_o,
\label{eq:R}
\end{equation}
while the capacitances are instead distributed proportional to the outlet area~\cite{Sankaran2012}.

\subsection{Agreement between low- and high-fidelity model outputs}

\begin{figure}[t]
\vspace{-6pt}
\centering
\includegraphics[width=0.9\textwidth, keepaspectratio]{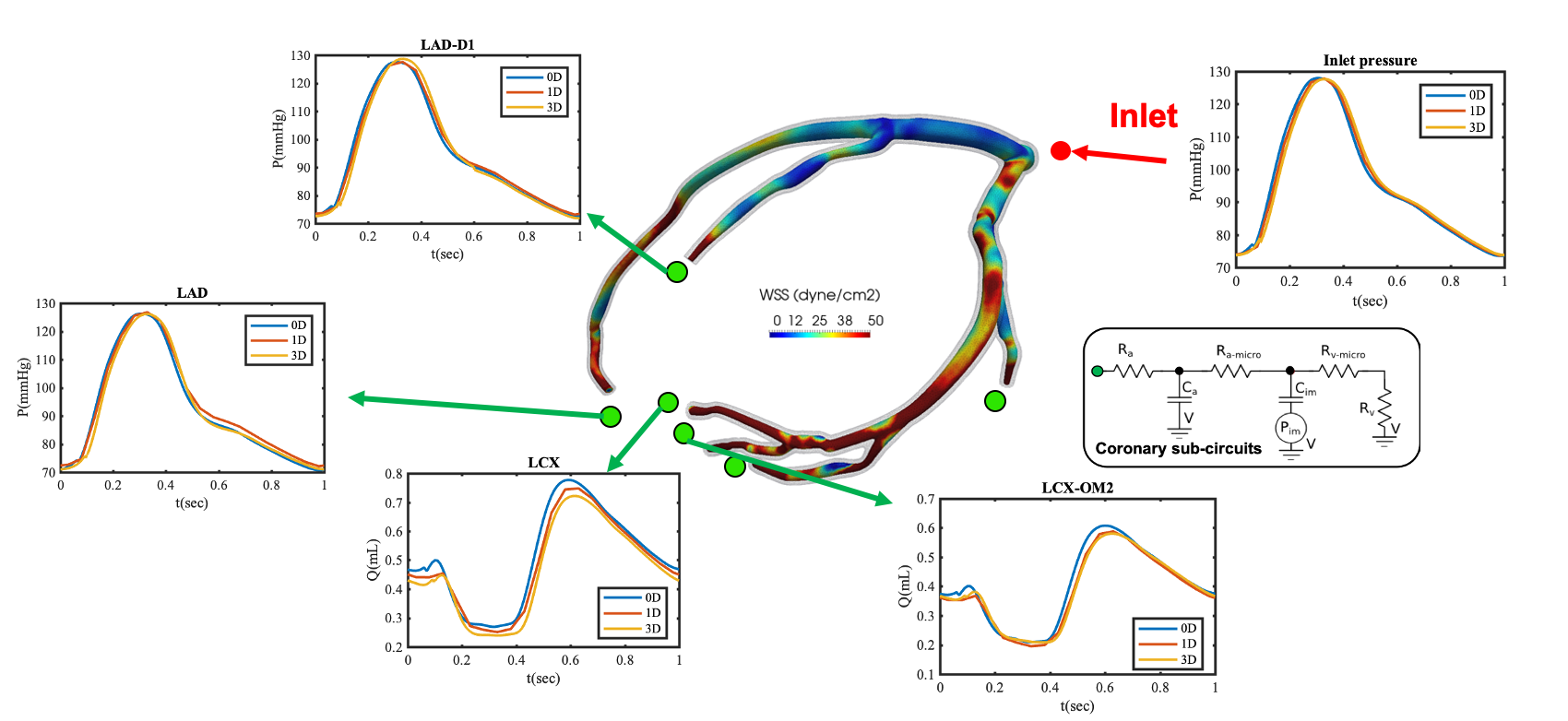}
\caption{Three-, one- and zero-dimensional model results with pulsatile inlet pressure and coronary outlet boundary condition.}\label{fig:3}
\end{figure}

Low fidelity one-dimensional models were extracted from the three-dimensional segmentation using a newly developed plugin in SimVascular~\cite{Lan2018} that automatically creates one-dimensional solver input files, containing the appropriate inlet and outlet boundary conditions. Similarly, low-fidelity zero-dimensional models were generated through a separate set of conversion tools from one-dimensional model segments. The agreement between low fidelity and three-dimensional model outputs was found to be excellent under both resistance-capacitance-resistance circuit (RCR) and coronary outlet boundary conditions as shown in Figure~\ref{fig:3} for selected flow and pressure quantities of interest.

\section{Multi-fidelity Monte Carlo uncertainty propagation}
\label{sec:MFUQ}

Consider a complete probability space $(\Omega, F, P)$ where $\Omega$ is a set of elementary events, $F$ is a Borel $\sigma$-algebra of $2^{\Omega}$, and $P$ a probability measure assuming values in [0,1] over events in $F$. 
Model inputs are represented though the random vector $\vec{\xi} = (\xi_1, \xi_2, ...\xi_d)$, with components $\xi_i:\Omega\rightarrow\Sigma_{i}, i=1,...,d$, ({{for $d$ dimensional random variable}}) having marginals $\xi_i \sim \rho_i (\xi_i)$ and joint probability density $\rho(\vec{\xi})$. Space and time are denoted by $\vec{x}\in\mathbb{R}^n$ and $t\in\mathbb{R}_{+}$, respectively.
In this study, we focus on the \emph{forward problem} in uncertainty quantification, $Q = Q(\vec{x}, \vec{\xi}, t)$, (through its mean $\mathop{\mathbb{E}}[Q]$ or variance $\mathop{\mathbb{V}}[Q]$), where, for a given realization $\vec{\xi}^{(i)}$ of the random inputs, $Q^{(i)} = Q(\vec{\xi}^{(i)})$ is determined through the numerical solution of a high-fidelity model. For simplicity, in what follows we assume space $\vec{x}$ and time $t$ to be fixed, leading to $Q = Q(\vec{\xi})$.

The Monte Carlo estimator $\hat{Q}^{\text{MC}}_N$ of the expected value of $Q$ based on $N$ realizations is defined as
\beq
\mathop{\mathbb{E}}[Q]=\int_{\boldsymbol{\Sigma}} Q(\vec{\xi})\rho(\vec{\xi})d\vec{\xi}  \simeq\hat{Q}^{\text{MC}}_{N}=\frac{1}{N}\sum_{i=1}^{N}Q^{(i)},\,\,\text{with variance}\,\,\mathop{\mathbb{V}}[\hat{Q}^{\text{MC}}_{N}]=\frac{\mathop{\mathbb{V}}[Q]}{N}.
\label{eq:MC}
\eeq
The variance $\mathop{\mathbb{V}}[\hat{Q}^{\text{MC}}_{N}]$ can be reduced by either increasing its denominator or reducing its numerator. The former can be achieved through \emph{stratified} or \emph{low-discrepancy} sampling sequences, such as Latin Hypercube or Quasi Monte Carlo sampling (QMC), which offer a more uniform coverage of the random input range $\boldsymbol{\Sigma}$, and have been shown to improve the convergence rate of MC estimators up to $\mathcal{O}(1/N^{2})$. In this study, we use Sobol'~\cite{Sobol1976} QMC sequences, as shown in Figure~\ref{fig:4}.

Alternatively, reduction in the numerator $\mathbb{V}[Q]$ can be achieved by introducing \emph{approximate control variate} Monte Carlo estimators, which combine contributions from two different model fidelities (low-fidelity LF and high-fidelity HF, respectively).
The Monte Carlo estimator for the high-fidelity model $\hat{Q}^{\text{HF}}$ is replaced by $\hat{Q}^{\text{CV,HF}}$ which embeds a correction based on the LF model~\cite{Pasupathy2012,Ng2014}
\begin{equation}
\hat{Q}^{\text{HF,CV}}_{N_{\text{HF}}}(r)=\hat{Q}^{\text{HF}}_{N_{\text{HF}}}+\alpha\,\left(\hat{Q}^{\text{LF}}_{N_{\text{HF}}}-\mathbb{E}[Q^{\text{LF}}]\right) \approx \hat{Q}^{\text{HF}}_{N_{\text{HF}}}+\alpha\,\left(\hat{Q}^{\text{LF}}_{N_{\text{HF}}}-\hat{Q}^{\text{LF}}_{N_{\text{LF}}}\right),
\end{equation}
where $N_{\text{LF}} = N_{\text{HF}} + \Delta_{\text{LF}} = N_{\text{HF}}\,(1 + r)$, and the additional LF realizations $\Delta_{\text{LF}} = r N_{\text{HF}}$ are used to estimate $\mathbb{E}[Q^{\text{LF}}]$ as $\hat{Q}^{\text{LF}}_{N_{\text{LF}}}$.
Similarly to the MC estimator, $Q^{\text{HF,CV}}$ is unbiased with variance
\begin{equation}
\mathop{\mathbb{V}}[\hat{Q}_{N_{\text{HF}}}^{\text{HF,CV}}]=\mathop{\mathbb{V}}[\hat{Q}_{N_{\text{HF}}}^{\text{HF}}]+\alpha^2 \mathop{\mathbb{V}}[\hat{Q}_{N_{\text{HF}}}^{\text{LF}}]+2\,\alpha\, \mathbb{C}[\hat{Q}_{N_{\text{HF}}}^{HF},\hat{Q}_{N_{\text{HF}}}^{LF}].
\label{eq:CV}
\end{equation}
Thus, the regression coefficient $\alpha$ that minimize~\eqref{eq:CV} can be determined as
\begin{equation}
\alpha=-\rho \sqrt{\frac{\mathop{\mathbb{V}}[\hat{Q}^{\text{HF}}_{N_{\text{HF}}}]}{\mathbb{V}[\hat{Q}^{\text{LF}}_{N_{\text{HF}}}]}},
\label{eq:CV2}
\end{equation}
where $\rho$ is Pearson's correlation coefficient between the LF and HF estimators. 
Substituting the optimal $\alpha$ from~\eqref{eq:CV2} into~\eqref{eq:CV}, leads to
\begin{equation}\label{equ:minVarMF}
\mathop{\mathbb{V}}[\hat{Q}^{\text{HF,CV}}_{N_{\text{HF}}}]=\mathop{\mathbb{V}}[\hat{Q}_{N_{\text{HF}}}^{\text{HF}}]\left(1-\frac{r}{1+r}\,\rho^2\right).
\end{equation}
Expression~\eqref{equ:minVarMF} suggests how a sufficiently large correlation between HF and LF models is the main determinant of variance reduction and, since $\rho^2\in(0, 1)$ approximate control variate estimators are always superior to their \emph{vanilla} Monte Carlo counterpart. 

\section{Input and output uncertainties}\label{sec:UQmodeling}

{ In an effort to focus our attention on relevant random inputs, essential for the physiological admissibility of the coronary outputs of interest, we consider two uncertain input parameters in the modeling of the coronary circulation model. The first random input parameter is the left coronary artery inlet pressure waveform imposed at the inlet of the left coronary model, { P$_\text{in}$} whose intra-patient variability is determined from repeated clinical data acquired \emph{in-vivo} through cardiac catheterization~\cite{Seo2020}. The second random input parameter is the time-derivative of the intramyocardial pressure in the LPN model for coronary circulation, P$_\text{im,t}$, imposed at the outlets of the model, whose variability is instead assumed. In our study we assumed these two parameters are independent and then conducted two separate propagation studies.}  

\subsection{Karhunen-Lo\`eve representation of the random inputs}\label{sec:coronaryPresUncertainty}

\noindent 

\begin{figure}[]
\vspace{-6pt}
\centering
\includegraphics[width=0.75\textwidth, keepaspectratio]{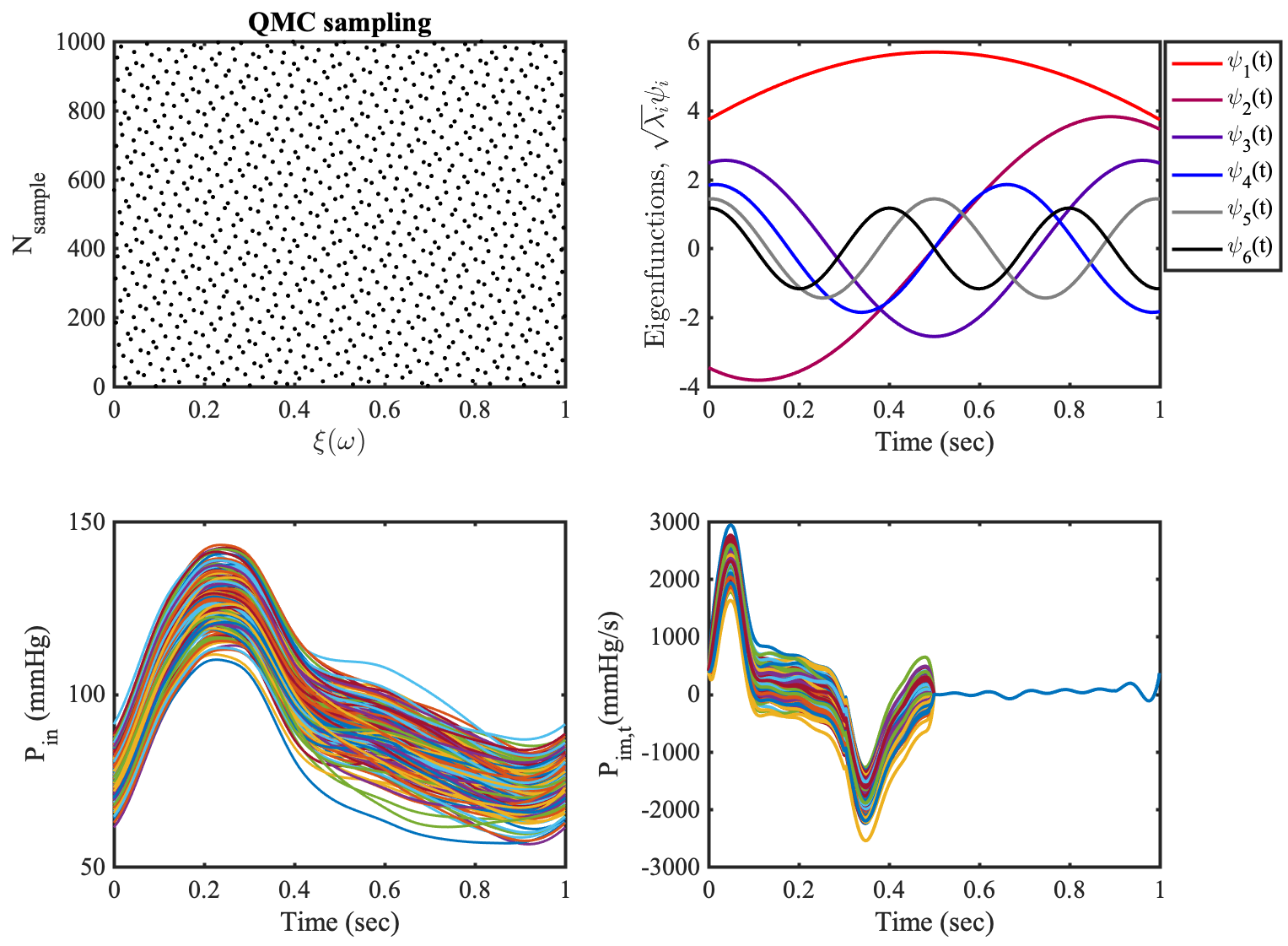}
\caption{Stochastic modeling of the random inputs. (a) 1000 samples from a Quasi Monte Carlo Sobol' sequence. (b) Six largest eigenfunctions of the exponential covariance kernel. (c) 200 inlet pressure realizations from the Karhunen-Lo\`eve (K-L) expansion~\eqref{eq:2}. (d) 200 realizations for the intramyocardial pressure time-derivative.}\label{fig:4}
\end{figure}

We approximate the left coronary pressure waveform, { $\text{P}_\text{in}$}, with a wide-sense stationary Gaussian process in time with exponential covariance, $\text{\bf K}(t,t')=\sigma^2\, \text{exp}(-|t-t'| / l_c)$, where $t$ and $t'$ are two arbitrary time points, $l_c$ is the correlation length, and $\sigma^2$ the process variance~\cite{ghanem2003stochastic}. Under these circumstances, for a sufficiently large truncation level { $N_\text{KL}$}, such a process admits the representation 
\begin{equation}\label{eq:2}
\text{P}(t,\omega)\approx\hat{\text{P}}(t)+\sum_{i=1}^{N_\text{KL}} \sqrt{\lambda_i}\,\psi_i(t)\,\xi_{i}(\omega),
\end{equation}
where $\vec{\xi}(\omega) =(\xi_1(\omega),\xi_2(\omega),\dots,\xi_N(\omega)),\,\omega\in\Omega$ is a collection of independent standard Gaussian random variables and $\lambda_i$, $\psi_i(t)$ are the eigenvalues and eigenvectors of the covariance kernel $\text{\bf K}(t,t')=\sum^{N_\text{KL}}_{i=1} \lambda_i \psi_i(t) \psi_i(t')$, respectively (\cite{Maitre2002}).
{ A satisfactory approximation of $\text{P}_\text{in}(t,\omega)$ is obtained using only the eigenmodes associated with the four largest eigenvalues ({{$N_\text{KL}$=4}}), when the correlation length is chosen to $l_c=\text{T}/2$=0.5 sec. In other words, { P$_\text{in}(t,\omega)$ is modeled with the K-L expansion with a four dimensional random variable.} We select $l_c=0.5$ based on the decay rate of the eigenvalue spectrum of the exponential covariance function. Specifically, the largest eigenvalue spectra compared to the first eigenvalue, $\lambda_i/\max_{i}(|\lambda_i|)$, are $\lambda_2/\lambda_1=34\%$, $\lambda_3/\lambda_1=14\%$, $\lambda_4/\lambda_1=7\%$, and $\lambda_i/\max_{i}(|\lambda_i|)$ decays to less than 5 percent after the first four modes ($i>4$). } For P$_\text{in}(t,\omega)$, the process standard deviation is set equal to 7\% of the mean, as per repeated pressure measurements from cardiac catheterization in six patients~\cite{Seo2020}. Samples from $\xi_{i}(\omega)$, $i=1,\dots,4$ are obtained by projecting the four-dimensional Sobol' sequence in Figure~\ref{fig:4}(a) through the inverse cumulative distribution function of a multivariate Gaussian. 
Results are illustrated in Figure~\ref{fig:4}. A collection of eigenfunctions $\psi_i(t),\,i=1,\dots,6$ from the selected covariance kernel is shown in Figure~\ref{fig:4}(b), while Figure~\ref{fig:4}(c) contains an ensemble of inlet pressure realizations from~\eqref{eq:2}.

Due to the significant challenge of measuring the intramyocardial pressure, { P$_\text{im}$}, \emph{in-vivo}, the left ventricular pressure and its time-derivative are often used as a viable replacement~\cite{Sankaran2012, Ramachandra2017, Tran2017, Tran2019}. 
This is clearly an approximation, whose effect on the simulation results is rarely discussed in the literature, despite the key importance of intramyocardial pressure in defining the diastolic character of coronary flow. This may relate to the use of open loop boundary conditions with prescribed coronary flow in many studies. 
Here, rather than a flow, we prescribe an inlet pressure and additionally model the uncertain intramyocardial pressure time-derivative P$_\text{im,t}$ as a stochastic process in time.
{ P$_\text{im,t}(t,\omega)$ is modeled with the K-L expansion with a four dimensional random variable.} We perturbed a baseline intramyocardial pressure time derivative~\cite{Tran2017} using a zero-mean Gaussian process having standard deviation equal to $\sigma=0.1\cdot\max_{t\in[0,T]}\,\text{P}_\text{im,t}$, with $T$ the heart cycle duration (Figure~\ref{fig:4}(d)). 

\subsection{Quantities of interests}\label{sec:UQmethods}
We focused on three hemodynamic quantities of interests (QoIs), flow rate and pressure at each outlet (Q$_i$, P$_i$, $i=1,...,n_o$), and branch wall shear stress (WSS). In more detail, the shear stress $\vec{\tau}({\bf x},t)$ was averaged over one heart cycle in time and over the vessel circumference in space so that
\beq
\text{WSS}_i\left(z\right)=\int_{C}\frac{1}{\text{T}}\left|\int_{0}^\text{T} \vec{\tau}({\bf x},t)\,dt\right| ds,
\eeq
where $z$ is the spatial distance along the vessel branch centerline, and $C$ is the lumen circumference at the branch cross section. Finally we averaged $\text{WSS}_i(z)$ along the centerlines of each vessel branch to obtain a single averaged value for WSS. For zero-dimensional models, the wall shear stress is determined based on Poiseuille flow conditions on an ideal vessel with circular cross section of radius $r$
\begin{equation}
|\tau| = \frac{4\,\mu\,Q}{\pi\,r^{3}},
\label{eq:twass}
\end{equation}
where $Q$ is the instantaneous flow rate, $\mu$ the dynamic viscosity and the vessel radii are based on a one-dimensional model anatomy.

\subsection{Computational cost}

Outputs for each model fidelity were computed using a fixed number of input parameter realizations. 
Specifically, we solved a { total} of 400 high-fidelity (HF) cardiovascular model instances in parallel using 48 cores { for each HF simulation}, either through the Comet cluster available through a XSEDE allocation, or using resources from the Center for Research Computing at the University of Notre Dame. All simulations were performed over four cardiac cycles, the first half consisting of QMC realizations of the inlet pressure $\text{P}_\text{in}$, while the remaining 200 with stochastic intramyocardial pressure $\text{P}_\text{im}$. 
One- and zero-dimensional models were solved 1,000 and 10,000 times, respectively for each random input.
Relative and cumulative simulation costs are reported in Tables~\ref{table:1} and ~\ref{table:2}, respectively.
\begin{table}[ht!]
\centering
\begin{tabular}{c ||c | c }
\hline
Solver& Cost & Effective Cost\\
& (1 Simulation) & (1 Simulation) \\
\hline
\hline
3D& 1775 hrs& 1\\
1D & 8.5 mins& 7.98$\times10^{-5}$\\
0D & 5 secs& 7.82$\times10^{-7}$\\
\hline
\end{tabular}
\caption{Absolute and relative simulation costs for all three model fidelity. { The simulation cost is reported in CPU times, in which computing times from all processors are summed.}}
\label{table:1}
\end{table}
\begin{table}[ht!]
\centering
\begin{tabular}{c ||c | c | c| c}
\hline
Method& Effective Cost & No. {3D} & No. {1D} & No. {0D}\\
& (Percentage to QMC) & simulations  & simulations & simulations \\
\hline
\hline
QMC& 100\%& 200 & 0 & 0 \\
MF-1D & 108.13\%& 200 & 1,000 & 0\\
MF-0D & 100.78\%& 200 & 0 & 10,000\\
\hline
\end{tabular}
\caption{Effective cost and associated number of model evaluations for high-fidelity Monte Carlo and multi-fidelity estimators.}
\label{table:2}
\end{table}

\subsection{Correlations of high- and low-fidelity models}\label{sec:UQdetails}

\begin{figure}[]
\vspace{-6pt}
\centering
\includegraphics[width=1.0\textwidth, keepaspectratio]{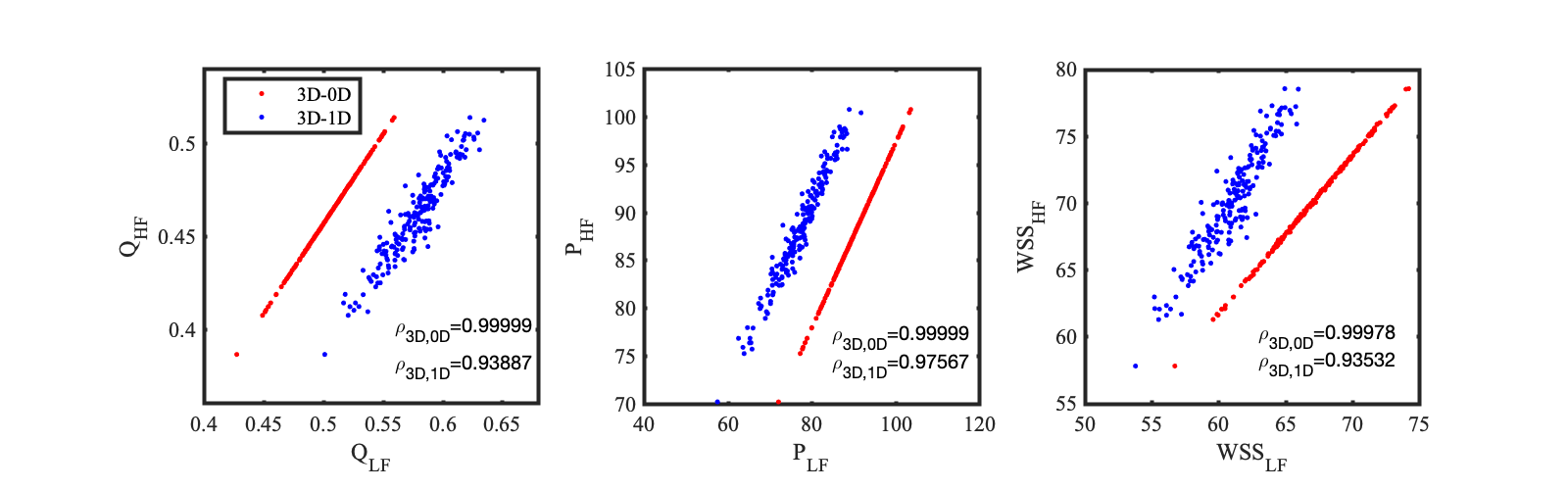}
\caption{Low- versus high-fidelity model outputs at LCx branch and Pearson correlations under intramyocardial pressure uncertainty.}\label{fig:5}
\end{figure}

Following the discussion on approximate control variate Monte Carlo estimators in Section~\ref{sec:MFUQ}, it should be clear how the correlations among high- and low-fidelity model outputs are the main determinant of variance reduction.
We observe very large average Pearson correlations ($\rho>0.9$) across all QoIs and all coronary branches, as reported in Tables~\ref{table:3} and ~\ref{table:4}.
Once again we would like to emphasize that a high correlation between model outputs does not necessarily imply a close agreement between high- and low-fidelity QoIs, as shown in Figure~\ref{fig:5}.
\begin{table}[ht!]
\centering
\begin{tabular}{c ||c c | c c  |c c }
\hline
QoI& \multicolumn{2}{ c |}{Pressure} & \multicolumn{2}{ c |}{Flow} & \multicolumn{2}{ c }{WSS}\\
\hline
Branch&3D-1D & 3D-0D&3D-1D & 3D-0D&3D-1D & 3D-0D\\
\hline
LCx&0.9756 &0.9999   &0.9388   &0.9999   &0.9353   &0.9997\\
LCx-OM$_1$&0.9756 &0.9999   &0.9281   &0.9999   &0.9092   &0.9998\\
LCx-OM$_2$&0.9753 &0.9999   &0.9360   &0.9999   &0.9292   &0.9997\\
LCx-OM$_3$&0.9758 &0.9999   &0.9315   &0.9999   &0.9234   &0.9999\\
LAD&0.9758 &0.9999   &0.9420   &0.9999   &0.9077   &0.9995\\
LAD-D$_1$&0.9716 &0.9959   &0.9295   &0.9999   &0.9040   &0.9998\\
\hline
\end{tabular}
\caption{High- to low-fidelity model correlations under inlet pressure waveform uncertainty. LCx - Left circumflex artery, OM - Obtuse marginal, LAD - Left anterior descending artery{, D - Diagonal branch of the LAD.} }\label{table:3}
\end{table}
\begin{table}[ht!]
\centering
\begin{tabular}{c ||c c | c c  |c c }
\hline
QoI& \multicolumn{2}{ c |}{Pressure} & \multicolumn{2}{ c |}{Flow} & \multicolumn{2}{ c }{WSS}\\
\hline
Branch&3D-1D & 3D-0D&3D-1D & 3D-0D&3D-1D & 3D-0D\\
\hline
LCX               &0.9886  &0.9960  &0.9857   &0.9999   & 0.9864   &0.9969\\
LCX-OM$_1$& 0.9909 &0.9942   &0.9911   &0.9999   &0.9913   &0.9984\\
LCX-OM$_2$& 0.9888 &0.9949   &0.9872   &0.9999   &0.9877   &0.9972\\
LCX-OM$_3$&0.9894 &0.9968   &0.9902   &0.9999   &0.9904   &0.9984\\
LAD               &0.9822 &0.9999  &0.9829    &0.9999   &0.9831   &0.9944\\
LAD-D$_1$   &0.9207 &0.9297  &0.9905   &0.9999   &0.9906   &0.9984\\
\hline
\end{tabular}
\caption{High- to low-fidelity model correlations under intramyocardial pressure uncertainty.}
\label{table:4}
\end{table}
%
\section{Uncertainty propagation results}\label{sec:results}
\subsection{Uncertainty in the inlet pressure waveform}\label{sec:uqPress}

\begin{figure}[ht!]
\vspace{-6pt}
\centering
\includegraphics[width=1.0\textwidth, keepaspectratio]{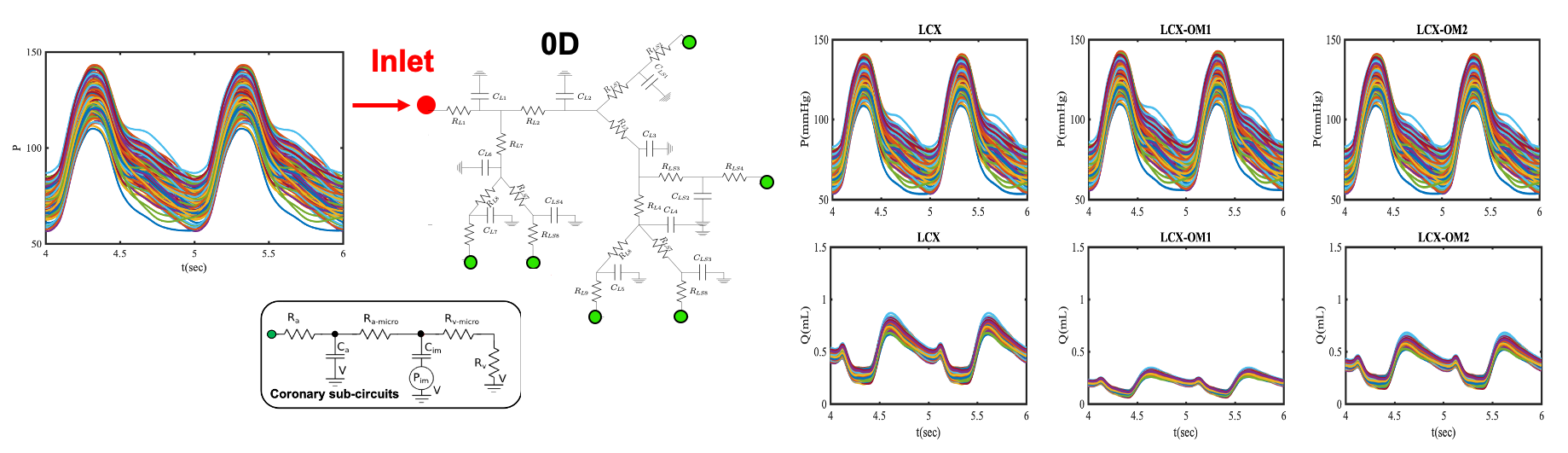}
\caption{Schematic workflow showing how the uncertainty in the inlet pressure waveform is propagated through a zero-dimensional left coronary artery model, specifically for outlet pressure and flow QoIs in LCx branches.}\label{fig:6}
\end{figure}
\begin{figure}[ht!]
\vspace{-6pt}
\centering
\includegraphics[width=1.0\textwidth, keepaspectratio]{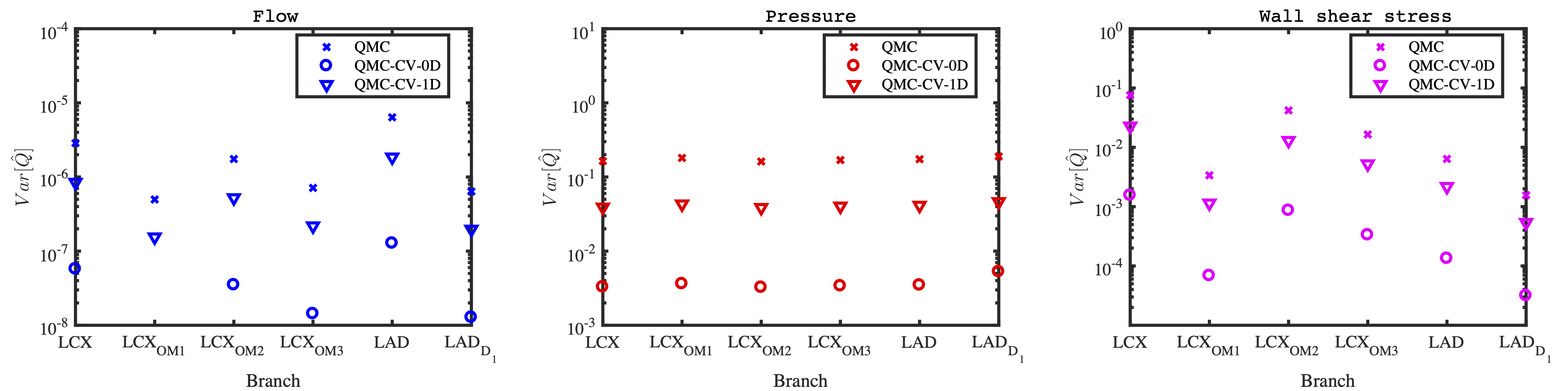}
\caption{Variance reduction for QMC and approximate control variate estimators.}\label{fig:7}
\end{figure}

To provide a preliminary idea of how the variability in inlet pressure waveform affects the model outputs, we first performed 200 realizations from the zero-dimensional model in Figure~\ref{fig:6}, and observed similar results as in previous studies from our research group~\cite{Seo2020}. 
Next, we compared the variance reduction obtained for QoIs in six left coronary artery branches resulting from {\it vanilla} Monte Carlo sampling and approximate control variate estimators with one- (CV-1D) and zero-dimensional (CV-0D) low-fidelity models, respectively (Figure~\ref{fig:7}).
From a pilot run consisting of 200, 1000, and 10000 model evaluations, the CV-1D estimator reduces the variance by a factor of three to four, while an order of magnitude reduction is achievable using the CV-0D estimator.

\begin{figure}[ht!]
\vspace{-6pt}
\centering
\includegraphics[width=0.6\textwidth, keepaspectratio]{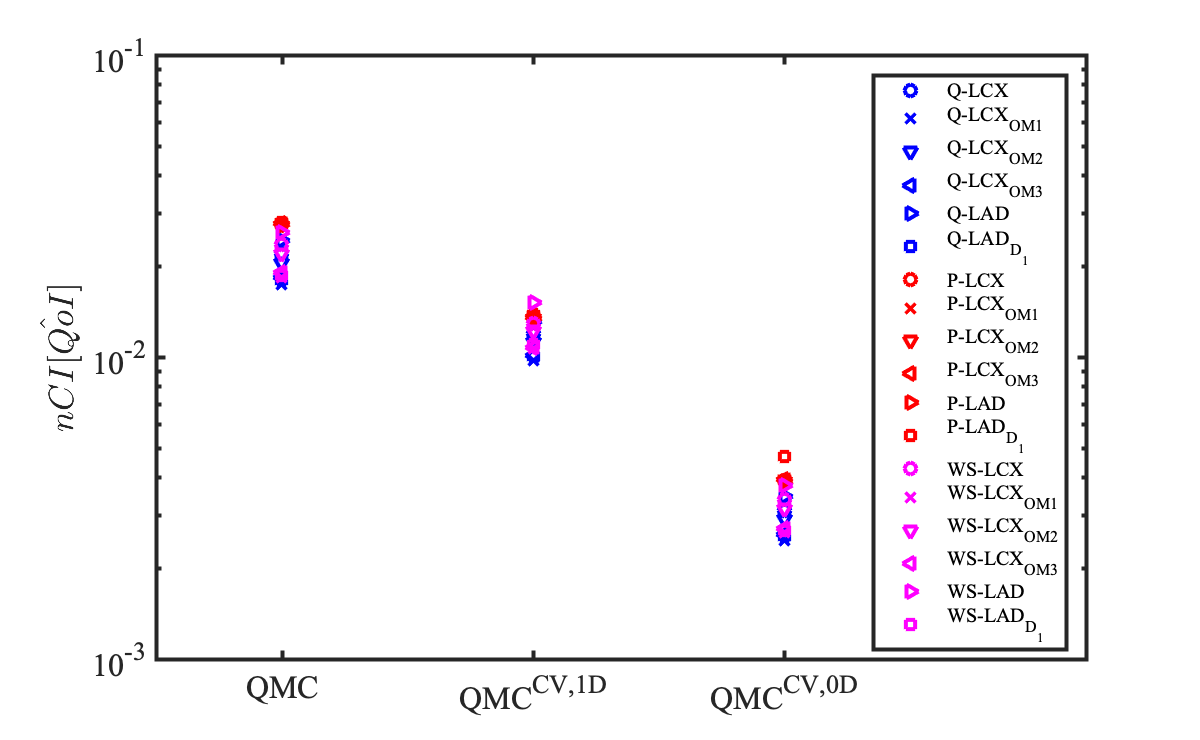}
\caption{Estimator the normalized confidence interval improvements from approximate control variate Monte Carlo.}\label{fig:8}
\end{figure}

{ We use the normalized confidence interval {  \emph{nCI}} of a generic estimator $\hat{Q}$ as six times the coefficient of variation in accordance with previous literature \cite{fleeter2019multilevel,Geraci2015}, 
\beq
{nCI}[\hat{Q}]= \frac{6\sqrt{\mathop{\mathbb{V}}[\hat{Q}]}}{\mathop{\mathbb{E}}[\hat{Q}]}=\frac{(\mu+3\sigma)-(\mu-3\sigma)}{\mu}=\frac{6\sigma}{\mu}. 
\eeq
The amplitude of $\pm 3\sigma$ (i.e., three standard deviations) confidence interval normalized by the mean corresponds to the 99.7$\%$ confidence interval normalized by the expected value. Smaller values of $nCI[\hat{Q}]$ indicates a higher level of confidence in the estimates from $\hat{Q}$.} The {  \emph{nCI}} of QMC, CV-1D and CV-0D estimators is shown in Figure~\ref{fig:8} for all relevant QoIs. Similar to previous results, reductions in {  \emph{nCI}} of one order of magnitude result from the CV-0D estimator. Finally, we note that the discussion above is equally valid for pressure/flow and for wall shear stress QoIs.

\subsection{Uncertainty in the intramyocardial pressure time-derivative}\label{sec:uqIntra}

\begin{figure}[ht!]
\vspace{-6pt}
\centering
\includegraphics[width=1.0\textwidth, keepaspectratio]{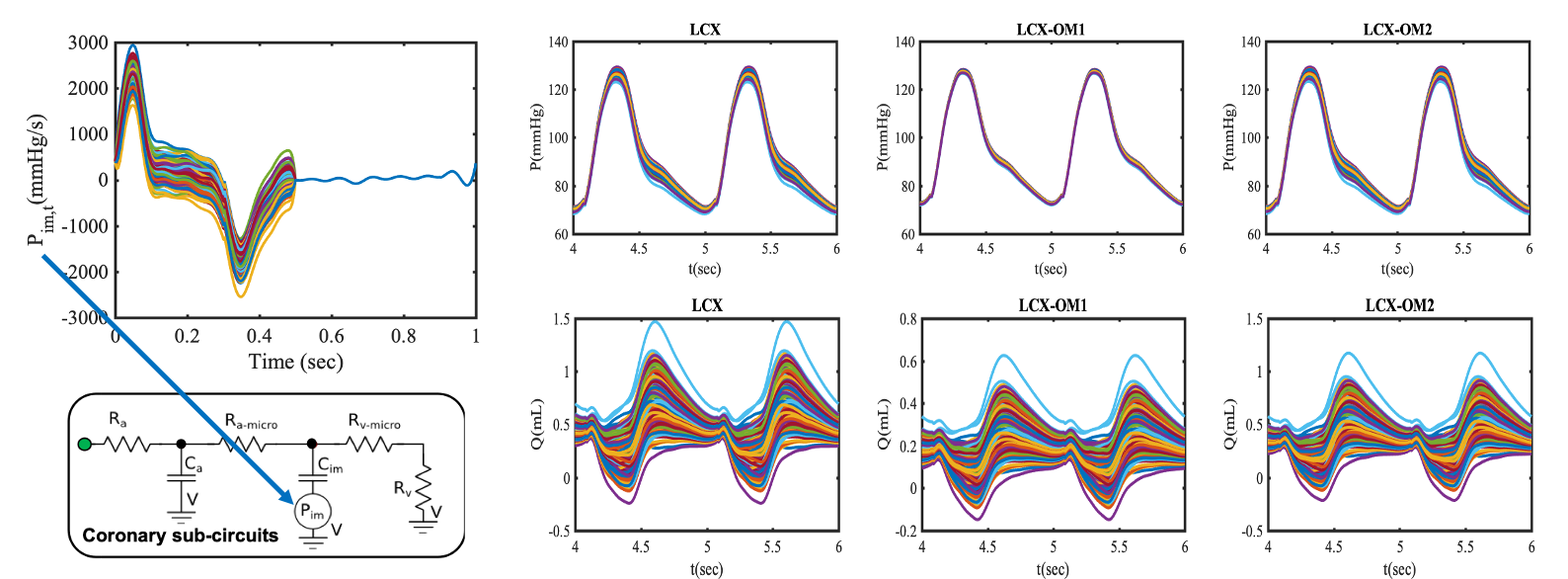}
\caption{Ensemble of model outputs for the zero-dimensional model of the left coronary circulation under intramyocardial pressure uncertainty.}\label{fig:9}
\end{figure}
\begin{figure}[ht!]
\vspace{-6pt}
\centering
\includegraphics[width=1.0\textwidth, keepaspectratio]{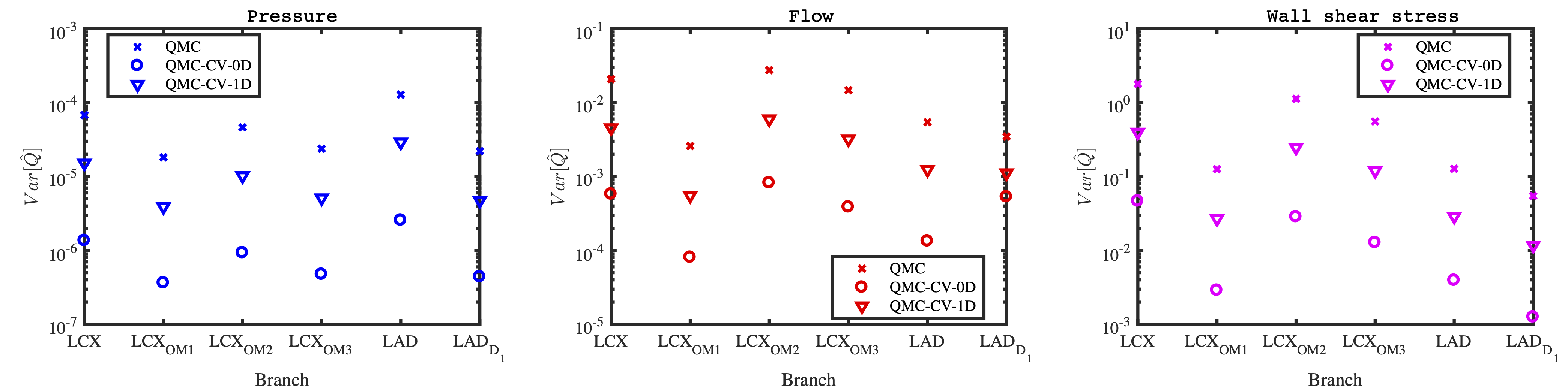}
\caption{Variance reduction from approximate control variate estimators under intramyocardial pressure uncertainty.}\label{fig:10}
\end{figure}
\begin{figure}[ht!]
\vspace{-6pt}
\centering
\includegraphics[width=0.6\textwidth, keepaspectratio]{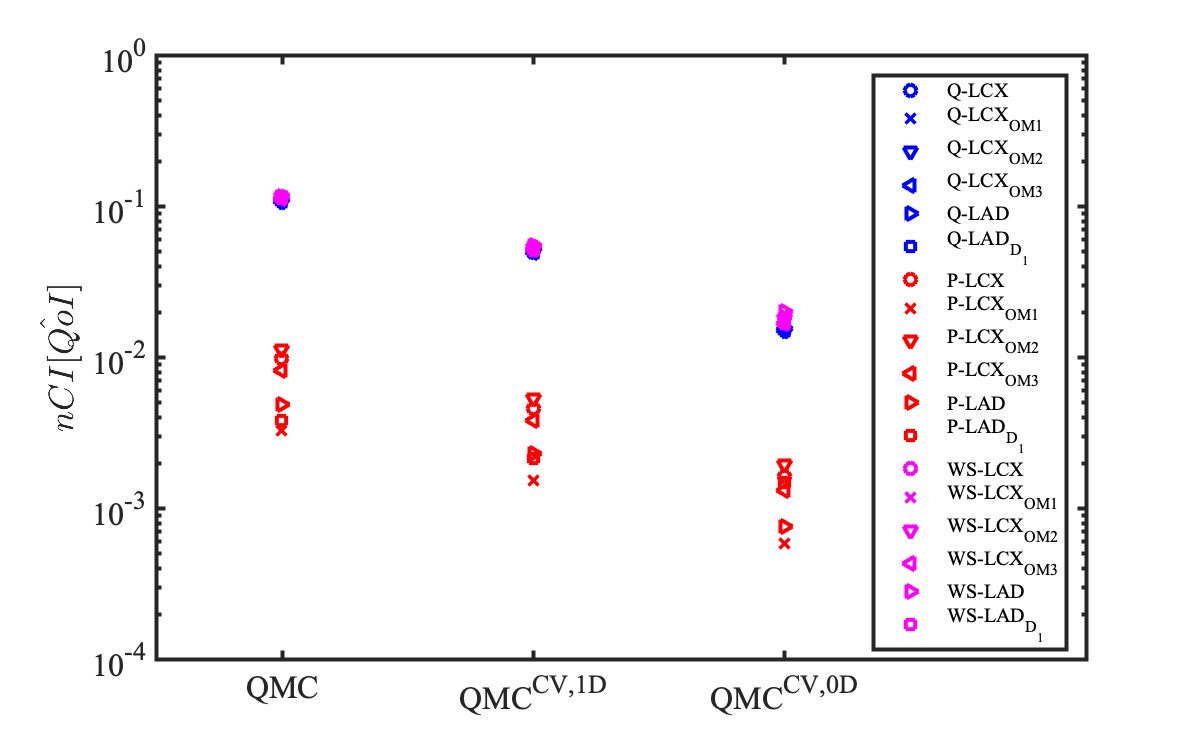}
\caption{Improvements in {  \emph{nCI}} from approximate control variate estimators under intramyocardial pressure rate uncertainty.}\label{fig:11}
\end{figure}

Variability in the intramyocardial pressure time-derivative significantly affects outlet flow and time-averaged wall shear stress uniformly across all branches ($\sim$30\% variation~\cite{Seo2020}), leaving the pressure almost unaltered, as shown in Figure~\ref{fig:9}.
Variance reduction for QMC and multifidelity estimators is quantified in Figure~\ref{fig:10}, while their {  \emph{nCI}} is shown in Figure~\ref{fig:11}.

Similarly to the findings in the previous section, both variance reduction and {  \emph{nCI}} are greatly improved by the multifidelity estimators with superior performance of CV-0D. This is due to the Pearson correlations being very high over all the QoIs for this latter estimator and the larger number of zero-dimensional simulations (10,000 for each input uncertainty) included in the pilot run. 
This suggests zero-dimensional models to be particularly well suited for low-fidelity or reduced-order representations of the coronary sub-circulation due to their almost negligible computational cost and highly correlated QoIs. 
Finally, we observe how outlet pressure variability due to intramyocardial pressure uncertainty is limited if compared to flow or wall shear stress QoIs. 
This justifies the clustering of {  \emph{nCI}} results in Figure~\ref{fig:11}.

We evaluate the performance of the multi-fidelity UQ using extrapolation of the computing cost in terms of the equivalent hi-fidelity model runs. We first set a range of targeted { \emph{nCI}} and estimate the computational cost to achieve the target { \emph{nCI}} by extrapolation using Equations \ref{eq:MC} and \ref{equ:minVarMF}. {{Specifically, the number of equivalent hi-fidelity runs, $N_{\text{eq,HF}}$, is computed as 
\beq 
N_\text{\text{eq,HF}}=\frac{1}{C_\text{HF}} (N_\text{HF} C_\text{HF} + N_\text{LF} C_\text{LF}),
\label{eq:eqNHF}
\eeq
where $C_\text{HF}$ and $C_\text{LF}$ are the computational cost and $N_\text{HF}$ and $N_\text{LF}$ are the number of simulations of hi- and low- fidelity models, respectively. We set a target \emph{nCI} of 0.2, 0.1, 0.05, 0.025, 0.01, 0.005, and extrapolate the cost of the simulation to reduce \emph{CI} using the variance reduction rule in equation\ref{eq:MC}, in which the variance is reduced by the inverse of the number of simulations}.}

\begin{figure}[ht!]
\vspace{-6pt}
\centering
\includegraphics[width=0.95\textwidth, keepaspectratio]{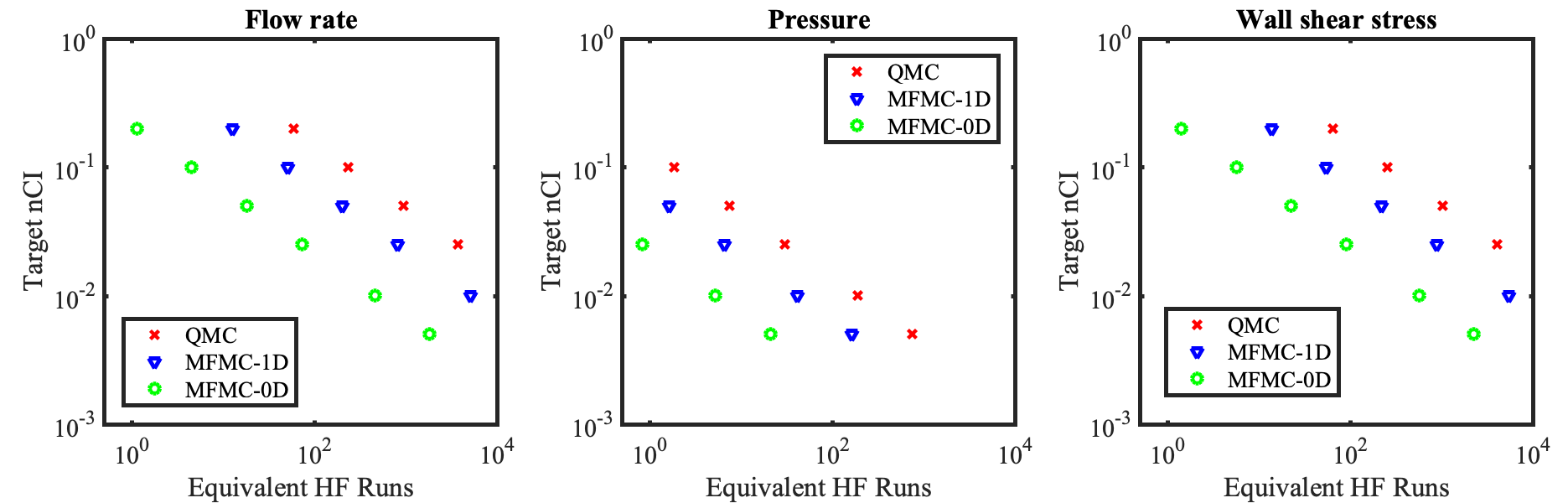}
\caption{Extrapolated equivalent high-fidelity model costs for target {  \emph{nCI}} for MC, MFMC with 1-D model, and MFMC with 0D model. QoI are at the LCX branch when the intramyocardial pressure is perturbed.}\label{fig:13}
\end{figure}

In Figure~\ref{fig:13}, we summarize the extrapolated computational cost to achieve a targeted {  \emph{nCI}} of the Monte Carlo estimator. The Multi-Fidelity Monte Carlo (MFMC) methods dramatically reduce the computational cost for providing accurate estimators of Monte-Carlo, especially for MF with the 0-D models. 

\section{Discussion}\label{sec:Discussion}

Recent advances in model-based diagnostics of coronary artery disease have demonstrated how models can be successfully integrated in clinical routines, but have raised questions related to their robustness, in light of possible uncertainty or ignorance associated with their input parameters. 
In this study, we focused on two clinically relevant parameters, the inlet pressure for coronary circulation sub-networks, and the intramyocardial pressure time-derivative.
Stochastic processes in time are used to inject uncertainty in these parameters, leading to a representation by a finite collection of independent random variables following Karhunen-Lo\`eve expansion. 
Inlet pressure waveform uncertainty is directly estimated from repeated measurements obtained during cardiac catheterization in six patients~\cite{Seo2020}, while variability in intramyocardial pressure is assumed equal to 10\% of its maximum systolic rate.

We have quantified the variance in model outputs expectations for a three-dimensional multi-scale cardiovascular model with deformable walls, ALE fluid-structure interaction and conforming fluid and wall mesh interfaces.
We also generated computationally inexpensive one- and zero-dimensional low-fidelity representations from three-dimensional segmentation, using a recently developed pipeline in SimVascular. Coronary boundary conditions were also implemented for all low-fidelity formulations and thoroughly validated against three-dimensional simulation results. 
Despite a common misconception that UQ analysis can easily become computationally intractable for large cardiovascular models containing millions of degrees of freedom, we were able to characterize confidence in expectations of clinically relevant quantities of interest under a fixed computational budget of approximately 200 high-fidelity model runs.

This is possible thanks to the high correlations between high- and low-fidelity outputs, even if these outputs are not necessarily in perfect agreement with each other, or in other words low-fidelity QoIs do not need in general to be close to their high-fidelity counterpart.
Our results highlight how CV-1D estimators of mean pressure, flow and wall shear stress have variance which is reduced by a factor of 3 to 4 with respect to QMC estimators, with accuracy improvements of the same order. CV-0D estimators show superior performance with improvements up to one order of magnitude and total simulation cost equal to less than 1 percent with respect to high-fidelity three-dimensional models. The computational cost savings using extrapolation of the current results shows that one can save one or two orders of magnitude compared to high-fidelity model evaluations alone to achieve a targeted {  \emph{nCI}} by using the multi-fidelity framework. 

Estimator variance and {  \emph{nCI}} are computed, in this study, from a pilot run consisting of a fixed number of simulations selected \emph{a-priori}. { A better approach would be to adaptively refine the number of simulations needed to achieve a certain variance reduction or to satisfy a certain computational budget from an small initial pilot run. 
This will be easily achieved through a software interface between SimVascular and the Dakota UQ software platform (developed at Sandia National Laboratory), currently under development~\cite{fleeter2019multilevel}.}

{ The high correlations between high- and low- fidelity models obtained in this study can be attributed to the fact that the current quantities of interest are spatially averaged global variables, and that the current cardiovascular model does not involve complex flows (e.g. flow recirculation or turbulence) or abrupt changes in the geometry (e.g. bifurcation, valve, or stenosis). We note that the 3D-1D approach cannot be replaced by 3D-0D when quantities of interests includes local variations of wall shear stress, atheroprone area, and wall deformations, as those are not obtainable from the 0D modeling.}

{ We remark that the MFMC approach provides many advantages over non-intrusive stochastic collocation methods, which were generally regarded as an accurate choice for uncertainty propagation. In our previous work \cite{Seo2020}, we demonstrated that convergence of QMC method provided similar performance against the stochastic collocation methods. However, MFMC can reduce variance by orders of magnitude at only a small fraction of the cost of one HF additional, while the next step refinement of stochastic collocation requires on the order of tens to hundreds HF simulations, since it requires a discrete increase in samples for refinements. The stochastic collocation method also has other shortcomings against the MC approach, for example, non-convergent behavior for discontinuous response surfaces.}

Future work will extend the present results for high-dimensional, arbitrary distributed and correlated random inputs. There are multiple sources of uncertainty in cardiovascular simulations, including boundary conditions, material properties and geometry that should be combined in future studies to offer a complete picture of uncertainty in predictions. The effects of these uncertainty sources used to be studied independently. In more realistic cases these uncertainty sources may be correlated. By reducing the computational cost, the multi-fidelity framework will be exploited to study the effect of multiple sources.

\acknowledgements

This work was supported by NIH grant (NIH NIBIB R01-EB018302), NSF SSI grants 1663671 and 1339824, and NSF CDS\&E CBET 1508794. This work used the Extreme Science and Engineering Discovery Environment (XSEDE)~\cite{XSEDE}, which is supported by National Science Foundation grant number ACI-1548562 and computational resources from the Centre for Research Computing at the University of Notre Dame. The authors would like to thank Aekaansh Verma, and Jonathan Pham for many the fruitful discussions on multifidelity estimators, LPN custom solvers and high performance computing resources. We also acknowledge support from the open source SimVascular project at www.simvascular.org.

\bibliographystyle{IJ4UQ_Bibliography_Style}
\bibliography{UQpapers}

\begin{thebibliography}{10}

\bibitem{AHA2018}
Benjamin, E., Virani, S., Callaway, C., Chamberlain, A., Chang, A., Cheng, S.,
  Chiuve, S., Cushman, M., Delling, F., Deo, R., , Heart disease and stroke
  statistics-2018 update: a report from the {A}merican {H}eart {A}ssociation.,
  {\em Circulation}, 137(12):e67, 2018.

\bibitem{koo2011diagnosis}
Koo, B., Erglis, A., Doh, J., Daniels, D., Jegere, S., Kim, H., Dunning, A.,
  DeFrance, T., Lansky, A., Leipsic, J., and Min, J., Diagnosis of
  ischemia-causing coronary stenoses by noninvasive fractional flow reserve
  computed from coronary computed tomographic angiograms: results from the
  prospective multicenter {DISCOVER-FLOW} (diagnosis of ischemia-causing
  stenoses obtained via noninvasive fractional flow reserve) study, {\em
  Journal of the American College of Cardiology}, 58(19):1989--1997, 2011.

\bibitem{min2012diagnostic}
Min, J., Leipsic, J., Pencina, M., Berman, D., Koo, B., Van~Mieghem, C.,
  Erglis, A., Lin, F., Dunning, A., Apruzzese, P., Budoff, M., J.H., C.,
  Jaffer, F., Leon, M., Malpeso, J., Mancini, G., Park, S., Schwartz, R., Shaw,
  L., and Mauri, L., Diagnostic accuracy of fractional flow reserve from
  anatomic {CT} angiography, {\em Jama}, 308(12):1237--1245, 2012.

\bibitem{norgaard2014diagnostic}
N{\o}rgaard, B., Leipsic, J., Gaur, S., Seneviratne, S., Ko, B., Ito, H.,
  Jensen, J., Mauri, L., De~Bruyne, B., Bezerra, H., Osawa, K., Marwan, M.,
  Naber, C., Erglis, A., Park, S., Christiansen, E., Kaltoft, A., Lassen, J.,
  B{\o}tker, H., and Achenbach, S., Diagnostic performance of noninvasive
  fractional flow reserve derived from coronary computed tomography angiography
  in suspected coronary artery disease: the {NXT} trial (analysis of coronary
  blood flow using ct angiography: Next steps), {\em Journal of the American
  College of Cardiology}, 63(12):1145--1155, 2014.

\bibitem{Marsden2014}
Marsden, A., Optimization in cardiovascular modeling, {\em Annual Review of
  Fluid Mechanics}, 46:519--546, 2014.

\bibitem{Marsden2015}
Marsden, A. and Esmaily-Moghadam, M., Multiscale modeling of cardiovascular
  flows for clinical decision support, {\em Applied Mechanics Reviews},
  67:030804, 2015.

\bibitem{Kung2011b}
Kung, E., Les, A., Figueroa, A., Medina, F., Arcautre, K., Wicker, R.,
  McConnell, M., and Taylor, C., In vitro validation of finite element analysis
  of blood flow in deformable models, {\em Annals of Biomedical Engineering},
  39(7):1947--1960, 2011.

\bibitem{Kung2014}
Kung, E., Kahn, A., Burns, J., and Marsden, A., In vitro validation of
  patient-specific hemodynamic simulations in coronary aneurysms caused by
  {K}awasaki disease, {\em Cardiovascular Engineering and Technology},
  5(2):189--201, 2014.

\bibitem{Steinman2012}
Steinman, D. and et~al., Variability of computational fluid dynamics solutions
  for pressure and flow in a giant aneurysm: the {ASME} 2012 {S}ummer
  {B}ioengineering {C}onference {CFD} challenge, {\em Journal of Biomechanical
  Engineering}, 135(021016):1--13, 2013.

\bibitem{taylor2013computational}
Taylor, C., Fonte, T., and Min, J., Computational fluid dynamics applied to
  cardiac computed tomography for noninvasive quantification of fractional flow
  reserve: scientific basis, {\em Journal of the American College of
  Cardiology}, 61(22):2233--2241, 2013.

\bibitem{yang2010constrained}
Yang, W., Feinstein, J., and Marsden, A., Constrained optimization of an
  idealized {Y}-shaped baffle for the {F}ontan surgery at rest and exercise,
  {\em Computer methods in applied mechanics and engineering},
  199(33-36):2135--2149, 2010.

\bibitem{Ramachandra2017}
Bangalore~Ramachandra, A., Kahn, A., and Marsden, A., Patient-specific
  simulations reveal significant differences in mechanical stimuli in venous
  and arterial coronary grafts, {\em Journal of cardiovascular translational
  research}, 9(4):279--290, 2016.

\bibitem{Noelia2017}
Grande~Gutierrez, N., Shirinsky, O., Gagarina, N., Lyskina, G., Fukazawa, R.,
  Ogawa, S., Burns, J., Marsden, A., and Kahn, A., Assessment of coronary
  artery aneurysms caused by {K}awasaki disease using transluminal attenuation
  gradient analysis of computerized tomography angiograms, {\em The American
  Journal of Cardiology}, 120:556--562, 2017.

\bibitem{Long2014}
Long, C., Marsden, A., and Bazilevs, Y., Shape optimization of pulsatile
  ventricular assist devices using {FSI} to minimize thrombotic risk, {\em
  Computional Mechanics}, 54:921--932, 2014.

\bibitem{Humphrey2008}
Humphrey, J.D. and Taylor, C., Intracranial and abdominal aortic aneurysms:
  Similarities, differences, and need for a new class of computational models,
  {\em Annu. Rev. Biomed. Eng.}, 10:221--246, 2008.

\bibitem{migliavacca2002mechanical}
Migliavacca, F., Petrini, L., Colombo, M., Auricchio, F., and Pietrabissa, R.,
  Mechanical behavior of coronary stents investigated through the finite
  element method, {\em Journal of Biomechanics}, 35(6):803--811, 2002.

\bibitem{Gundert2012}
Gundert, T., Marsden, A., Yang, W., Marks, D., and LaDisa, J.J., Identification
  of hemodynamically optimal coronary stent designs based on vessel caliber,
  {\em IEEE Transactions on Biomedical Engineering}, 58(7):1992--2002, 2012.

\bibitem{Figueroa2006}
Figueroa, C., Vignon-Clementel, I., Jansen, K., Hughes, T., and Taylor, C., A
  coupled momentum method for modeling blood flow in three-dimensional
  deformable arteries, {\em Computer Methods in Applied Mechanics and
  Engineering}, 195(41-43):5685--5706, 2006.

\bibitem{Esmaily2013res}
Esmaily-Moghadam, M., Hsia, T., and Marsden, A., A non-discrete method for
  computation of residence time in fluid mechanics simulations, {\em Physics of
  Fluids}, 25(11):110802, 2013.

\bibitem{Tran2017}
Tran, J., Schiavazzi, D., Bangalore~Ramachandra, A., Kahn, A., and Marsden, A.,
  Automated tuning for parameter identification and uncertainty quantification
  in multi-scale coronary simulations, {\em Computers \& Fluids}, 142:128--138,
  2017.

\bibitem{Schiavazzi2017}
Schiavazzi, D., Baretta, A., Pennati, G., Hsia, T., and Marsden, A.,
  Patient-specific parameter estimation in single-ventricle lumped circulation
  models under uncertainty, {\em International journal for numerical methods in
  biomedical engineering}, 33(3):e02799, 2017.

\bibitem{Xiu2007}
Xiu, D. and Sherwin, S., Parametric uncertainty analysis of pulse wave
  propagation in a model of a human arterial network, {\em Journal of
  Computational Physics}, 226:1385--1407, 2007.

\bibitem{Brault2016}
Brault, A., Dumas, L., and Lucor, D., Uncertainty quantification of inflow
  boundary condition and proximal arterial stiffness--coupled effect on pulse
  wave propagation in a vascular network, {\em International Journal for
  Numerical Methods in Biomedical Engineering}, 33(e2859), 2017.

\bibitem{Tran2019}
Tran, J., Schiavazzi, D., Kahn, A., and Marsden, A., Uncertainty quantification
  of simulated biomechanical stimuli in coronary artery bypass grafts, {\em
  Computer Methods in Applied Mechanics and Engineering}, 345:402--428, 2019.

\bibitem{yin2019one}
Yin, M., Yazdani, A., and Karniadakis, G., One-dimensional modeling of
  fractional flow reserve in coronary artery disease: Uncertainty
  quantification and {B}ayesian optimization, {\em Computer Methods in Applied
  Mechanics and Engineering}, 353:66--85, 2019.

\bibitem{Chen2013}
Chen, P., Quarteroni, A., and Rozza, G., Simulation-based uncertainty
  quantification of human arterial network hemodynamics, {\em International
  Journal for Numerical Methods in Biomedical Engineering}, 29:698--721, 2013.

\bibitem{xiu2002wiener}
Xiu, D. and Karniadakis, G., The {W}iener--{A}skey polynomial chaos for
  stochastic differential equations, {\em SIAM journal on scientific
  computing}, 24(2):619--644, 2002.

\bibitem{wan2005adaptive}
Wan, X. and Karniadakis, G., An adaptive multi-element generalized polynomial
  chaos method for stochastic differential equations, {\em Journal of
  Computational Physics}, 209(2):617--642, 2005.

\bibitem{le2004multi}
Le~Ma{\i}tre, O., Najm, H., Ghanem, R., and Knio, O., Multi-resolution analysis
  of wiener-type uncertainty propagation schemes, {\em Journal of Computational
  Physics}, 197(2):502--531, 2004.

\bibitem{schiavazzi2014sparse}
Schiavazzi, D., Doostan, A., and Iaccarino, G., Sparse multiresolution
  regression for uncertainty propagation, {\em International Journal for
  Uncertainty Quantification}, 4(4), 2014.

\bibitem{schiavazzi2017generalized}
Schiavazzi, D., Doostan, A., Iaccarino, G., and Marsden, A., A generalized
  multi-resolution expansion for uncertainty propagation with application to
  cardiovascular modeling, {\em Computer methods in applied mechanics and
  engineering}, 314:196--221, 2017.

\bibitem{witteveen2012simplex}
Witteveen, J. and Iaccarino, G., Simplex stochastic collocation with random
  sampling and extrapolation for nonhypercube probability spaces, {\em SIAM
  Journal on Scientific Computing}, 34(2):A814--A838, 2012.

\bibitem{doostan2011non}
Doostan, A. and Owhadi, H., A non-adapted sparse approximation of {PDEs} with
  stochastic inputs, {\em Journal of Computational Physics}, 230(8):3015--3034,
  2011.

\bibitem{blatman2011adaptive}
Blatman, G. and Sudret, B., Adaptive sparse polynomial chaos expansion based on
  least angle regression, {\em Journal of Computational Physics},
  230(6):2345--2367, 2011.

\bibitem{peherstorfer2018survey}
Peherstorfer, B., Willcox, K., and Gunzburger, M., Survey of multifidelity
  methods in uncertainty propagation, inference, and optimization, {\em Siam
  Review}, 60(3):550--591, 2018.

\bibitem{gorodetsky2018generalized}
Gorodetsky, A., Geraci, G., Eldred, M., and Jakeman, J., A generalized
  framework for approximate control variates, {\em arXiv preprint
  arXiv:1811.04988}, 2018.

\bibitem{fleeter2019multilevel}
Fleeter, C., Geraci, G., Schiavazzi, D., Kahn, A., and Marsden, A., Multilevel
  and multifidelity uncertainty quantification for cardiovascular hemodynamics,
  {\em Comput. Methods Appl. Mech. Engrg.}, 365, 2020.

\bibitem{kim2010patient}
Kim, H., Vignon-Clementel, I., Coogan, J., Figueroa, C., Jansen, K., and
  Taylor, C., Patient-specific modeling of blood flow and pressure in human
  coronary arteries, {\em Annals of biomedical engineering}, 38(10):3195--3209,
  2010.

\bibitem{sankaran2012patient}
Sankaran, S., Moghadam, M., Kahn, A., Tseng, E., Guccione, J., and Marsden, A.,
  Patient-specific multiscale modeling of blood flow for coronary artery bypass
  graft surgery, {\em Annals of biomedical engineering}, 40(10):2228--2242,
  2012.

\bibitem{Updergrove2016}
Updegrove, A., Wilson, N., Merkow, J., Lan, H., Marsden, A., and Shadden, S.,
  Sim{V}ascular: an open source pipeline for cardiovascular simulation, {\em
  Annals of Biomedical Engineering}, 45(3):525--541, 2016.

\bibitem{Seo2019}
Seo, J., Schiavazzi, D., and Marsden, A., Performance of preconditioned
  iterative linear solvers for cardiovascular simulations in rigid and
  deformable vessels, {\em Computational Mechanics},
  doi.org/10.1007/s00466-019-01678-3, 2019.

\bibitem{Bazilevs2007}
Bazilevs, Y., Calo, V., Cottrell, J., Hughes, T., Reali, A., and Scovazzi, G.,
  Variational multiscale residual-based turbulence modeling for large eddy
  simulation of incompressible flows, {\em Computer Methods in Applied
  Mechanics and Engineering}, 197:173--201, 2007.

\bibitem{Esmaily2015}
Esmaily-Moghadam, M., Bazilevs, Y., and Marsden, A., A bi-partitioned iterative
  algorithm for solving linear systems arising from incompressible flow
  problems, {\em Computer Methods in Applied Mechanics and Engineering},
  286:40--62, 2015.

\bibitem{Trilinos}
Heroux, M., Bartlett, R., Howle, V., Hoekstra, R., Hu, J., Kolda, T., Lehoucq,
  R., Long, K., Pawlowski, R., Phipps, E., Salinger, A., Thornquist, H.,
  Tuminaro, R., Willenbring, J., Williams, A., and Stanley, K., An overview of
  the {T}rilinos project, {\em ACM Transactions on Mathematical Software},
  31(3):397--423, 2005.

\bibitem{Seo2020}
Seo, J., Schiavazzi, D.E., and Marsden, A.L., The effects of clinically-derived
  parametric data uncertainty in patient-specific coronary simulations with
  deformable walls, {\em arXiv preprint arXiv:1908.07522}, 2020.

\bibitem{Hughes1973}
Hughes, T. and Lubliner, J., On the one-dimensional theory of blood flow in the
  larger vessels, {\em Mathematical Biosciences}, 18:161--170, 1973.

\bibitem{Wan2002}
Wan, J., Steele, B., Spicer, S.A., Strohband, S., Feijo, G.R., Hughes, T.J.,
  and Taylor, C.A., A one-dimensional finite element method for
  simulation-based medical planning for cardiovascular disease, {\em Comput
  Methods Biomech Biomed Engin}, 5:195--206, 2002.

\bibitem{Vignon2004}
Vignon, I. and Taylor, C., Outflow boundary conditions for one-dimensional
  finite element modeling of blood flow and pressure waves in arteries,
  39:361--374, 2004.

\bibitem{Vignon2006}
Vignon, I., A coupled multidomain method for computational modeling of blood
  flow, PhD thesis, Stanford University, 2006.

\bibitem{Milisic2004}
Milisic, V. and Quarteroni, A., Analysis of lumped parameter models for blood
  flow simulations and their relation with {1D} models, {\em ESAIM:
  Mathematical Modelling and Numerical Analysis}, 38:613--632, 2004.

\bibitem{Kim2009}
Kim, H., Vignon-Clementel, I., Figueroa, C., LaDisa, J., Jansen, K., Feinstein,
  J., and Taylor, C., On coupling a lumped parameter heart model and a
  three-dimensional finite element aorta model, {\em Annals of Biomedical
  Engineering}, 37(11):2153--2169, 2009.

\bibitem{Sankaran2012}
Sankaran, S., Esmaily-Moghadam, M., Kahn, A., Tseng, E., Guccione, J., and
  Marsden, A., Patient-specific multiscale modeling of blood flow for coronary
  artery bypass graft surgery, {\em Annals of Biomedical Engineering},
  40(10):2228--2242, 2012.

\bibitem{Esmaily2012}
Esmaily~Moghadam, M., Vignon-Clementel, I., Figliola, R., and Marsden, A., A
  modular numerical method for implicit 0{D}/3{D} coupling in cardiovascular
  finite element simulations, {\em Journal of Computational Physics},
  244:63--79, 2013.

\bibitem{Bogren1989}
Bogren, H., Klipstein, R., Firmin, D., Mohiaddin, R., Underwood, S., Rees, R.,
  and Longmore, D., Quantitation of antegrade and retrograde blood flow in the
  human aorta by magnetic resonance velocity mapping, {\em American Heart
  Journal}, 117(6):1214--1222, 1989.

\bibitem{Zhou1999}
Zhou, Y., Kassab, G., and Molloi, S., On the design of the coronary arterial
  tree: a generalization of {M}urray's law, {\em Physics in Medicine \&
  Biology}, 44(12):2929, 1999.

\bibitem{Changizi2000}
Changizi, M. and Cherniak, C., Modeling the large-scale geometry of human
  coronary arteries, {\em Canadian Journal of Physiology and Pharmacology},
  78:603--611, 2000.

\bibitem{Lan2018}
Lan, H., Updegrove, A., Wilson, N., Maher, G., Shadden, S., and Marsden, A., A
  re-engineered software interface and workflow for the open-source
  {S}im{V}ascular cardiovascular modeling package, {\em Journal of
  Biomechanical Engineering}, 140(024501), 2018.

\bibitem{Sobol1976}
Sobol, I., Uniformly distributed sequences with an additional uniform
  property., {\em USSR Comput. Math. Math. Phys.}, 16:236--242, 1976.

\bibitem{Pasupathy2012}
Pasupathy, R., Schmeiser, B.W., Taaffe, M.R., and Wang, J., Control-variate
  estimation using estimated control means., {\em IIE Trans}, 1044:381--385,
  200812.

\bibitem{Ng2014}
Ng, L. and Willcox, K., Multifidelity approaches for optimization under
  uncertainty., {\em Int J Numer Methods Eng}, 100:746--772, 2014.

\bibitem{ghanem2003stochastic}
Ghanem, R. and Spanos, P., {\em Stochastic finite elements: a spectral
  approach}, Courier Corporation, 2003.

\bibitem{Maitre2002}
Le~Maitre, O., Reagan, M., Najm, H., Ghanem, R., and Knio, O., A atochastic
  projection method for fluid flow {II}. random process, {\em Journal of
  Computational Physics}, 181:9--44, 2002.

\bibitem{Geraci2015}
Geraci, G., Eldred, M., and Iaccarino, G., A multifidelity control variate
  approach for the multilevel {M}onte {C}arlo technique, Tech.~Rep., Stanford
  University, 2015.

\bibitem{XSEDE}
Towns, J., Cockerill, T., Dahan, M., Foster, I., Gaither, K., Grimshaw, A.,
  Hazlewood, V., Lathrop, S., Lifka, D., Peterson, G., Roskies, R., Scott, J.,
  and Wilkins-Diehr, N., {XSEDE}: Accelerating scientific discovery, {\em
  Computing in Science and Engineering}, 16(5):62--74, 2014.

\end{thebibliography}
\end{document}